\documentclass[12pt,aps,final,eqsecnum,notitlepage, onecolumn,superscriptaddress,showpacs,
centertags,nobibnotes,nofootinbib]{revtex4}

\usepackage{stmaryrd,amsmath,amssymb,graphics,graphicx,bm,bbm,bbold}
\usepackage{amscd,amsbsy,array}
\usepackage{dsfont}

\begin{document}

\title{Galilean Conformal and Superconformal Symmetries}

\author{\firstname{Jerzy}~\surname{Lukierski}}
%%\email[] {lukier@ift.uni.wroc.pl}
\affiliation{Institute for Theoretical Physics,  University of Wroc{\l}aw, \\ pl. Maxa Borna 9, 50-204 Wroc{\l}aw,\ {e-mail:lukier@ift.uni.wroc.pl}}

%\date{\today}

\begin{abstract}
Firstly we discuss briefly three different algebras named as nonrelativistic (NR) conformal: Schr\"{o}dinger, Galilean conformal and  infinite algebra of local NR conformal isometries. Further we shall consider in some detail Galilean conformal algebra (GCA) obtained in the limit $c \to \infty$ 
from relativistic conformal algebra $O(d+1,2)$ ($d$ - number of space dimensions). Two different  contraction limits providing  GCA and some recently considered realizations will be briefly discussed. Finally by considering NR contraction of D=4 superconformal algebra the Galilei conformal superalgebra (GCSA) is obtained, in the formulation using complex Weyl supercharges.
\end{abstract}
\maketitle

%\tableofcontents

\section{Introduction}
The notion of nonrelativistic conformal symmetries was used at least in three-fold sense:
\\
\textit{a) Schr\"odinger symmetries.}

The $d$ dimensional Galilei group $G(d)$ can be represented as the following semidirect product
\begin{equation}\label{mosluke1}
G(d) = (O(d) \oplus R) \niplus T^{2d} \,,
\end{equation}
where $R$ describes the time translations (generator $H$), $O(d)$ - the $d$-dimensional space rotations (generators $J_{ij}= - J_{ji}$) and the Abelian subgroups $T^{2d}$ represents the space translations (generators $P_i$) and Galilean boosts (generator $B_i$).
Almost forty years ago \cite{mosluk1}--\cite{mosluk4}  to the corresponding Galilean algebra ${\mathfrak{g}}(d)$ there were added two generators $D$ and $K$, generating the dilatations (scale transformations) and the expansions
 (conformal transformations of time). New generators are forming together with $H$ the $d=0$ conformed algebra $O(2,1)$:
\begin{equation}\label{mosluke2}
[D,H]= - H\,,
\quad
[K,H]=-2D\,, 
\quad
[D,K] = K \,.
\end{equation}

In such a way we obtain the Schr\"odinger group 
${\textit{Schr}}(d)$, which is obtained from (\ref{mosluke1}) by the enlargement of 
$R\simeq O(1,1)$ to $O(2,1)$ group 
\begin{equation}\label{mosluke3}
{\textit{Schr}}(d) = (SO(d)\oplus SO(2,1))\niplus T^{2d}\,.
\end{equation}
The symmetries (\ref{mosluke3}) can be introduced as the set of transformations preserving the Schr\"odinger equation for free nonrelativistic massive particle (we put $\hbar=1$)
\begin{equation}\label{mosluke4}
\widehat{S}_m (d) \, \psi = 0 \,, \qquad
\widehat{S}_m (d) \equiv i  \frac{\partial}{\partial t} - \frac{\Delta^{(d)}}{2m}
\,.
\end{equation}
The parameter $m$ can be interpreted in geometric way as the central extension of Schr\"odinger algebra,
 which defines the
  ``quantum'' Schr\"odinger symmetries. The ``quantum'' Schr\"odinger group can be obtained also as the enlargement by dilatations and expansions of the
   centrally extended Galilean group, called ``quantum'' Galilean or Bargmann group.

\textit{b) Galilean conformal symmetries} \cite{mosluk5}--\cite{mosluk13}.

For $d\neq 1$ the Galilean conformal algebra is finite-dimensional and obtained by nonrelativistic contraction $c\to \infty$ of $D$-dimensional ($D=d+1$) relativistic conformal algebra $O(d+1,2)$. To two generators $D$, $K$ which enlarge  Galilei to Schr\"odinger algebra one adds $d$ Abelian generators $F_i$, describing Galilean accelerations defined by the  nonrelativistic limit of conformal translations. The structure of $d$-dimensional Galilean conformal algebra 
${\mathfrak{c}}(d)$, denoted as GCA, is described by the algebra with the following semi-direct product structure
\begin{equation}\label{mosluke5}
{\mathfrak{c}}(d)=(O(d)\oplus O(2,1)) \inplus T^{3d} \,,
\end{equation}
where $T^{3d}=(P_i,B_i, F_i)$. It should be stressed that

\textit{i)} From the contraction procedure follows that similarly as the pair of Galilean and Poincar\'{e} algebras, the Galilean and relativistic conformal algebras have the same dimension. 

\textit{ii)} GCA does not permit to introduce the mass-like parameter as its central extension, i.e.   it  describes only the symmetries of massless NR dynamical systems\footnote{In principle however one can introduce the GCA - invariant model with continuous spectrum of nonrelativistic masses.}

\textit{iii)} By putting $F_i=0$ in (\ref{mosluke5}) we do not arrive at the  algebra (\ref{mosluke3}), because in 
 Schr\"odinger and Galilean conformal algebras the boosts $B_i$ transforms differently under dilatations:
\begin{equation}\label{mosluke6}
{\textit{Schr}}(d): \ [D,B_i]=-B_i 
\qquad {\mathfrak{c}}(d): [D,B_i]=0 \,.
\end{equation}
The subalgebra of ${\mathfrak{c}}(d)$ obtained by putting $F_i =0$ as well as
${\textit{Schr}}(d)$ belong to the one-parameter family of generalized Schr\"odinger algebras $
{\textit{Schr}}_{z}(d)$ with the following commutator of generators $D$ and $B_i$~\footnote{The generalized Schr\"odinger algebras $ 
{\textit{Schr}}_{z}(d)$ were introduced by Henkel \cite{mosluk8}. If we still introduce arbitrary parameter defining the scaling properties of $H$ ($[D,H]=(y-1)H$) one obtains the  Milne-conformal algebra, with independent scaling of time and space (see also \cite{mosluk13}).}
\begin{equation}\label{mosluke7}
[D,B_i]=(1-z)B_i \,,
\end{equation}
where $z$ is called a dynamical exponent and 
${\textit{Schr}}(d)=
{\textit{Schr}}_2(d)(z=2)$, ${\mathfrak{c}}(d)\supset 
{\textit{Schr}}_1(d) (z=1)$.
The dilatations (scale transformations) described by 
${\textit{Schr}}_z(d)$ transform the nonrelativistic space-time coordinates ($t, \vec{x}=(x_1 \ldots x_d))$ as follows 
\begin{equation}\label{mosluke8}
t' = \lambda^z \, t \qquad 
\vec{x}{\, }' = \lambda \vec{x} \,.
\end{equation}
It is easy to see that the invariance of free Schr\"odinger equation (\ref{mosluke4}) is achieved if $z=2$.

\textbf{c)} Infinite-dimensional conformal extension of GCA \cite{mosluk7,mosluk14,mosluk8,mosluk9,mosluk15,mosluk12,mosluk13}.

The degeneracy of nonrelativistic metric in space-time allows to introduce infinite-dimensional Galilean conformal isometries described by vector fields satisfying  NR conformal Killing equations \cite{mosluk14,mosluk13}.
Finite-dimensional GCA (\ref{mosluke5}) is spanned by the vector fields
\begin{equation}\label{mosluke9}
X_{({\mathfrak{c}}(d))}=
\left(
\frac{1}{2} k t^2 + \lambda t +\varepsilon
 \right)
 \frac{\partial}{\partial t}
+
\left(
\omega_{ij} x_j + \lambda x_i + k  t x_i
- \frac{1}{2} \alpha_i t^2 + \beta_i t + \gamma_i
\right)
\frac{\partial}{\partial x_i} \,.
\end{equation}

Among infinite-dimensional Galilean conformal isometries it is distinguished the following set of NR Killing vector fields,
denoted in \cite{mosluk13} by ${\mathfrak{ cgal}}_z$, 
 generating reparametrization of time $t$, 
 local time-dependent NR space-time dilatations and translations ($t \to \xi (t), x_i \to 
 (1+ \frac{1}{z}\xi'(t))x_i
  + \eta_i (t)$)
as well as time-dependent $O(d)$ rotations
\begin{equation}\label{mosluke10}
X= \xi (t)\frac{\partial}{\partial t}
+
\left(
\omega_{ij} (t) x_j 
+  \frac{1}{z}\,\xi ' (t) x_i 
+ \eta_i (t)
\right)
\frac{\partial}{\partial x_i} \,.
\end{equation}
Formula (\ref{mosluke10}) contains as a subset the Galilean-conformal vector fields (\ref{mosluke9}) as well as the $z$-dependent Schr\"odinger algebra 
${\textit{Schr}}_z(d)$.

It should be added that the infinite-dimensional NR conformal isometries do not have their infinite-dimensional relativistic counterpart except if $d=1$.

Further we shall restrict our considerations to the finite-dimensional GCA (see (\ref{mosluke5}) and (\ref{mosluke9})).
In Sect.~2 we shall discuss the derivation of GCA and will consider two 
 versions of NR contraction procedure leading to GCA. 
 Subsequently we shall provide some 
 recent results about GLA realizations.
In Sect.~3 we shall describe the supersymmetrization of $D=4$ GCA
 \cite{mosluk16}--\cite{mosluk18}. In order to distinguish the presentation here from the one given in \cite{mosluk16} with real Majorana supercharges, we will formulate the Galilean conformal superalgebra in terms of complex Weyl supercharges. 
 In Sect.~4 we provide a brief outlook.

\section{Galilean Conformal Symmetries}

Following the derivation of Galilei algebra by NR contraction $c\to \infty$  of Poincar\'{e} algebra, one can as well perform the contraction $c \to \infty$ of $\frac{(d+2)(d+3)}{2}$ generators of relativistic conformal algebra $O(d+1,2)$. 
Denoting by $P_\mu = (P_0,P_i)$,
 $M_{\mu\nu} = (M_{ij} = \frac{1}{2} \epsilon_{ijk} M_k)$,
  $M_{i0}= N_i$) ($\mu,\nu=0,1, \ldots d; \ i,j=1,2\ldots d$)
  the Poincar\'{e} generators, we get $D=(d+1)$-dimensional relativistic conformal algebra by adding the dilatation generator $D$
  and conformal generators $K_ \mu=(K_0, K_i)$. The ``natural'' NR contraction $c\to \infty$ leads to Galilean conformal algebra 
  ${\mathfrak{c}}(d)= (H, P_i, M_k, B_k,D, K, F_i)$ if we rescale the relativistic generators as follows \cite{mosluk10}
  \begin{eqnarray}\label{mosluke2.1}
 & P_0 = \frac{H}{c} \,, \qquad \quad &
  N_i = cB_i \,,
  \cr
  & K_0 = cK  \,,
  \qquad \quad &
  K_i= c^2 F_i \,,
  \end{eqnarray}
  ($P_i, M_i, D$ remain unchanged). 
  One obtains besides the $O(2,1)$ algebra (see (\ref{mosluke2}))
  and $O(d)$ generators $M_{ij}$ the following
 $O(2,1) \otimes O(d)$ covariance relations ($i,j,k=1 \ldots d$)
 \begin{equation}\label{mosluke2.2}
 \begin{array}{lll}
 [H,P_i]= 0 \,,\qquad \qquad
 &[H,B_i]= P_i \,,
 \qquad \qquad
 &
 [H,F_i]=2B_i\,,
 \cr
 [K,P_i]= -2B_i  \,,\qquad \qquad
 & [K,B_i]= F_i \,,
 \qquad \qquad
 &
 [K,F_i]=0 \,,
 \cr
 [D,P_i]= -P_i  \,,\qquad \qquad
& [D,B_i]= 0\,,
 \qquad \qquad
 &
 [D,F_i]=F_i\,,
 \end{array}
 \end{equation}
 and ($l=1,2,3$)
 \begin{equation}\label{mosluk2.3}
 [M_{ij}, A_{k;l}] = \delta_{jk}\, A_{i;l} -
 \delta_{ik} \, A_{j;l} \,,
 \end{equation}
 where $A_{i;l}=(P_i,B_i,F_i$) describe the maximal Abelian subalgebra $T^{3d}$ (see (\ref{mosluke5})).
 
 We shall consider  the various contraction procedures leading to GCA and present some results on GCA realizations.
 
 \textbf{a)} Nonuniqueness of contraction procedure.
 
 Besides the ``natural'' or ``physical'' contraction limit (\ref{mosluke2.1}) one can use as well other contraction $O(d+1,2)
 {\xrightarrow[\lambda\to\infty]{\phantom{sssslhh}}}
   {\mathfrak{c}}(d)$
 if we observe that the semidirect structure of ${\mathfrak{c}}(d)$ given by (\ref{mosluke5}) can be obtained  by the canonical contraction
  from the following coset decomposition of 
  the relativistic conformal group
 \begin{equation}\label{mosluke2.4}
 O(d+1,2) = \mathbb{H} \cdot \mathbb{K} \,,
 \end{equation}
 where $\mathbb{H}=O(d)\otimes O(2,1)$ and
 $\mathbb{K}=\frac{O(d+1,2)}{O(d)\otimes O(2,1)}$.
 The corresponding generators of relativistic conformal algebra  ${\mathfrak{c}}_{rel}(d) = \widehat{h}\oplus \widehat{k}$
  $(\widehat{h}= \{h_i \}$,  $\widehat{k}=\{k_r \})$, describe (pseudo)Riemannian symmetric pair. If we perform the rescaling 
 \begin{equation}\label{mosluke2.5}
 h'_i = h_i \qquad \qquad k'_r = \lambda \, k_r
 \end{equation}
 we get in the limit $\lambda \to \infty$ exactly the algebraic structure
  (\ref{mosluke5}), with the $O(2,1)$ generators described by the relativistic conformal generators $P_0$, $K_0$ and $D$ \cite{mosluk11,mosluk17}. 
  The rescaling (\ref{mosluke2.5}) in comparison with (\ref{mosluke2.1}) is simpler, but the contraction parameter $\lambda$ can not be related in universal way (for all indices $r$ in (\ref{mosluke2.5})) with light velocity $c$. 
  In order to find the relation of two contraction procedures let us observe that the choice of generators describing $O(d+1,2)$ conformal algebra is not unique. 
  In particular one can introduce the following new basis
  \begin{equation}\label{mosluke2.6}
  \widetilde{P}_\mu = \widetilde{\lambda} \, P_\mu
  \qquad \qquad
  \widetilde{K}_{\mu} 
  = \widetilde{\lambda}^{-1} \, K_\mu \,,
  \end{equation}
  with generators $M_{\mu\nu}$ and $D$ unchanged
   ($\widetilde{M}_{\mu\nu}=M_{\mu\nu}$, $\widetilde{D}=D$). It is easy to check that composing the rescalings (\ref{mosluke2.1}) and
   (\ref{mosluke2.6}) one obtains
   \begin{equation}\label{mosluke2.7}
   \begin{array}{ll}
   \widetilde{P}_0 = \frac{\widetilde{\lambda}}{c}\, H \,,
   \qquad \qquad \quad
   & 
   \widetilde{K}_0 = \frac{c}{\widetilde{\lambda}} \, K\,,
   \cr  
   \widetilde{P}_i = c P_i\,,
   \qquad \qquad \quad
   & 
   \widetilde{K}_i = \frac{c^2}{\widetilde{\lambda}}\, F_i\,,
   \end{array}
   \end{equation}
   and
   \begin{equation}\label{mosluke2.8}
   \widetilde{N}_i = cB_i\,,
   \qquad \widetilde{M}_i = M_i \,,\qquad
   \widetilde{D} = \widetilde{D} \,.
   \end{equation}
 If we put $\widetilde{\lambda}=c$ the relations (\ref{mosluke2.7}--\ref{mosluke2.8}) become identical to (\ref{mosluke2.5})  provided that $\lambda  =c$.
 It should be added that in such a contraction scheme the dimensional analysis is consistent only if we identify the dimensionality of space and time, what implies that the rescaling parameter $c=\lambda =\widetilde{\lambda}$ should be considered as  dimensionless.
 
  The rescaling (\ref{mosluke2.6}) is the only one which preserves the relativistic conformal algebra.
  For GCA the class of rescalings preserving the algebraic structure is larger. One can show the invariance of 
  ${\mathfrak{c}}(d)$ under the following rescaling  depending on the  parameter $k$
 \begin{eqnarray}\label{mosluke2.9}
 &H' = \rho^{-k-1} \, H \,,
 \qquad \qquad
 K'=\rho^{k+1} \, K\,,
 \cr
 &P'_i = \rho^{-k} \, P_i\,,
 \qquad \qquad
 B'_i = \rho B _i \,,
 \qquad \qquad
 F'_i = \rho^{k+2} \, F_i\,,
 \cr
 &M'_{ij} = M_{ij} \,,
 \qquad \qquad
  D' = D \,.
 \end{eqnarray}
 If $k=0$ in such a way one gets the renormalization ($c'=\rho c$) of the rescaling (\ref{mosluke2.1}), and if $k=-1$ we obtain the modification ($\lambda'=\rho \lambda$) of the contraction parameter in (\ref{mosluke2.5}).
 
 \textbf{b)} The representations of GCA.
 
 The contractions (\ref{mosluke2.1}) and
 (\ref{mosluke2.5}) can be also applied to the space-time realization of relativistic conformal algebra\footnote{We consider here the spinless realization of relativistic conformal algebra.}
 \begin{eqnarray}\label{mosluke2.10}
&P_\mu = - \frac{\partial}{\partial x^\mu}\,,
\qquad \qquad &
K_\mu = x^2 
\frac{\partial}{\partial x^\mu}
- 2 x_\mu \, x_\nu
\frac{\partial}{\partial x^\nu}\,,
\cr\cr
&M_{\mu\nu} = x_\mu 
\frac{\partial}{\partial x^\nu}
- x_\nu \frac{\partial}{\partial x^\mu}\,,&
\qquad \qquad
D= x_\mu \frac{\partial}{\partial x^\mu}\,.
 \end{eqnarray}
 After substituting  $x_0=ct$ and performing the rescaling (\ref{mosluke2.1}) one gets the vector fields realization of GCA in nonrelativistic space-time $(x_i,t)$
 \begin{eqnarray}\label{mosluke2.11}
&
H =  \frac{\partial}{\partial t}\,,
\qquad\qquad
K = t^2
\frac{\partial}{\partial t}
- 2 t x_i \, 
\frac{\partial}{\partial x_i}\,,&
\cr\cr
&
P_{i} = -
\frac{\partial}{\partial x_i}\,,
\qquad \qquad
B_i=- t \frac{\partial}{\partial x_i}\,,
\qquad \qquad
F_i= -t^2\frac{\partial}{\partial x_i} \,,&
\cr\cr 
&
M_{ij}=x_i \frac{\partial}{\partial x_j}
- x_j  \frac{\partial}{\partial x_i}\,,
\qquad\qquad
D = x_i  \frac{\partial}{\partial x_i}
- t  \frac{\partial}{\partial t}\,.&
 \end{eqnarray}
 
 In order to obtain the rescaling (\ref{mosluke2.5}) one should rescale in (\ref{mosluke2.10})
    the space coordinates $x_i \to x'_i = \lambda x_i$
 and remain time coordinate unchanged. 
 Further, if we introduce in (\ref{mosluke2.11}) the rescaling of nonrelativistic space-time
 \begin{equation}\label{mosluke2.12}
 x'_i = \rho^k \, x_i\,,
 \qquad \qquad
 t'= \rho^{k+1} \, t \,,
 \end{equation}
 we obtain the rescaling (\ref{mosluke2.9}) of GCA.
 
 One can introduce as well the finite-dimensional spinors of GCA defined as the spinorial representations 
  of the compact part $O(d)\oplus O(2,1)$  of GCA (for that purpose we should consider the double coverings $\overline{SO(d)} \otimes SU(1,1)$). Further we shall consider $d=3$ and use
  $\overline{SO(3)}=SU(2)$.
  
  The fundamental spinor representation of $d=3$ GCA which can be called  the nonrelativistic $d=3$ twistor is identified with the double spinor $t_{\alpha;A}\in C^4$ $(\alpha=1,2; A=1,2)$ transforming as follows under $SU(2)\otimes SU(1,1)$ \cite{mosluk19}
  \begin{equation}\label{mosluke2.13}
  {\mathbb{t}}' = A \,{\mathbb{t}} \, B^T
   \qquad A \in SU(2) \quad B \in SU(1,1)\,,
  \end{equation}
  where $\mathbb{t} = \{t_{\alpha;A} \}$ is a $2\times 2$ complex matrix. The
  general nonrelativistic conformal spinors of rank ($n,m$) are described by the set of $2^{n+m}$ complex variables transforming as follows
  \begin{equation}\label{mosluke2.14}
  t'_{\alpha_1 \ldots \alpha_n; A_1 \ldots A_m}
  = A_{\alpha_1}^{\quad \alpha'_1} \ldots
  A_{\alpha_n}^{\quad \alpha'_n} \cdot
  t_{\alpha'_1 \ldots \alpha'_n; A'_{1} \ldots A'_{m}}
  \cdot
  (B^T)^{A'_1}_{\ \ A_1} \ldots
   (B^T)^{A'_m}_{\ \  A_m} \,.
  \end{equation}
  The fundamental ($n=m=1$) NR conformal spinors (twistors) leave invariant the following $SU(2) \times SU(1,1)$ norm
  \begin{equation}\label{mosluke2.15}
  \left\langle 
  \mathbb{t}^{(1)}, \mathbb{t}^{(2)}
  \right\rangle
  = \bar{t}^{(1)}_{\dot{\alpha}; \dot{A}}
  (\sigma_3)_{\dot{A} A} \, t^{(2)}_{\alpha;A} \,,
  \end{equation}
  where $\bar{t}^{(1)}_{\dot{\alpha}; \dot{A}}=(t_{\alpha;A})^\star$, and Pauli matrix $\sigma_3$ describes the $U(1,1)$ metric.
  
  An interesting question arises whether on NR level one can repeat the constructions known from the  Penrose formalism of relativistic twistors (see e.g. \cite{mosluk20,mosluk21}) which are defined as fundamental representations of  $SU(2,2)=SO(4,2)$.
  In such geometric framework twistor components are primary, and space-time  as well as relativistic phase space coordinates are composite. 
  First step in searching  of such NR counterpart is to look for NR twistor realizations of GCA. Because GCA has nonsemisimple structure, it contains Abelian subalgebra (see $T^{3d}$ in (\ref{mosluke5})), and that implies the following results \cite{mosluk19}.
  
  \textbf{i)} The NR twistorial realizations of GCA exist only in the space of $N$ twistors with $N\geq2$.

 \textbf{ii)}  Contrary to the twistor realizations of relativistic conformal algebra
 the NR twistor realizations of GCA  for any $N$ 
  do not provide the positive-definite Hamiltonian $H$.
 
 Other  GCA  realizations which lead to GCA-covariant classical mechanics models is provided by the application 
 of the geometric techniques of nonlinear realizations \cite{mosluk22}
 to various cosets of  Galilean conformal group.
 By using the inverse Higgs method \cite{mosluk23,mosluk24} applied to the Cartan-Maurer one-forms spanned by  GCA one gets the extension of standard conformal mechanics of de Alfaro, Fubini and Furlan \cite{mosluk25} in the presence of nonvanishing space dimensions as well as the geometric derivation in $D=2+1$ of the GCA model presented in~\cite{mosluk26b} and its generalization\footnote{The standard conformal mechanics \cite{mosluk25} employs as symmetry 
 $c(0)=O(2,1)$, with only conformal transformations of the time coordinate.}.
 
 Finally let us recall that dynamical realizations of GCA were obtained \cite{mosluk13} in the model describing Souriau ``Galilean photons'' \cite{mosluk26} and as well in the magnetic-like nonrelativistic limit of Maxwell electrodynamics \cite{mosluk27}.
 
 \section{Extended D=4 Relativistic Conformal Superalgebra  SU(2,2,N) and its Nonrelativistic Contraction}

 The $D=4$ $N$-extended relativistic conformal algebra \cite{mosluk28,mosluk29} is supersymmetrized by supplementing the 
 $N$ complex Weyl Poincar\'{e} supercharges
 $Q_{\alpha i} \bar{Q}_{\dot{\alpha}}^{\ i} =(Q_{\alpha i})^\star$ satisfying the relations\footnote{In the relation (\ref{mosluke3.2}) we did put the Poincar\'{e} superalgebra central charges equal to zero. It can be further shown that this requirement follows from the Jacobi identities for the extended conformal superalgebra \cite{mosluk28}.}
 \begin{eqnarray}\label{mosluke3.1}
 \{Q_{\alpha i}, \bar{Q}_{\beta}^{\ j} \}
 & = & 
  2(\sigma_\mu)_{\alpha \dot{\beta}} \, P^\mu  \,
  \delta_{i}^{\ j}\,,
  \\[12pt]
  \label{mosluke3.2}
   \{Q_{\alpha i}, {Q}_{\beta  j} \}
 & = & 
   \{\bar{Q}_{\dot{\alpha}}^{\ i}, \bar{Q}_{\dot{\beta}}^{\ j} \}=0 \,,
 \end{eqnarray}
 by $N$ complex conformal Weyl supercharges $S^i_\alpha, \bar{S}^{i}_{\dot{\alpha}}= (S_{\alpha i})^\star$, representing  ``supersymmetric roots'' of conformal momenta $K_\mu$
 \begin{eqnarray}\label{mosluke3.3}
 \{ S_{\alpha i}, \bar{S}_{\dot{\beta}}^{\ j} \}
  & = & 
  - 2(\sigma_\mu)_{\alpha \dot{\beta}} \, K^\mu  \,
  \delta_{i}^{\ j}\,,
  \\[12pt]
 \label{mosluke3.4}
  \{ S_{\alpha i}, {S}_{{\beta} j} \}
  & = & 
   \{ \bar{S}_{\dot{\alpha}}^{\ i},
    \bar{S}_{\dot{\beta}}^{\ j} \} = 0 \,.
 \end{eqnarray}
 Besides we have the relations
 \begin{subequations}
 \begin{eqnarray}\label{mosluke3.5}
  \{ Q_{\alpha i}, {S}_{{\beta} j} \}
  &=&
  \delta_{ij}
  [(\sigma_{\mu\nu})_{\alpha\beta}
  \, M^{\mu\nu} - 2 \epsilon_{\alpha \beta}
  (D+iA)] 
  - 2i \epsilon_{\alpha \beta} \, T_{ij} \,,
  \label{mosluke3.5a}
  \\[12pt]
  \label{mosluke3.5b}
    \{ \bar{Q}_{\dot{\alpha}}^{\ i}, 
    \bar{S}_{\beta}^{\ j} \}
  &=&
  \delta^{ij}
[(\bar{\sigma}_{\mu\nu})_{\dot{\alpha}\dot{\beta}}
  \, M^{\mu\nu} 
  - 2 \epsilon_{\dot{\alpha} \dot{\beta}}
  (D-iA)] 
  + 2i \epsilon_{\dot{\alpha} \dot{\beta}} \, 
  \bar{T}^{ij} \,,
  \\[12pt]
  \label{mosluke3.5c}
   \{ {Q}_{{\alpha} i}, 
    \bar{S}_{\dot{\beta}}^{\ j} \}
    &=&
     \{ \bar{Q}_{\dot{\alpha}}^{\ i}, 
    {S}_{{\beta} j} \} = 0\,,
    \\[12pt]
     \{ {S}_{{\alpha} i}, 
    {S}_{{\beta} j} \}
    &=&
     \{ \bar{S}_{\dot{\alpha}}^{\ i}, 
    \bar{S}_{\dot{\beta}}^{\ j} \} = 0\,,
 \label{mosluke3.5d}
 \end{eqnarray}
 \end{subequations}
 where $(T_{ij})^\star = \bar{T}^{ij}$,
 \begin{equation}\label{mosluke3.6}
 (\sigma_{\mu\nu})_{\alpha\beta}
 = \frac{i}{2}
  (\sigma_{[\mu} )_{\alpha \dot{\gamma}}
   (\bar{\sigma}_{\nu ]} )_{\dot{\gamma}\beta}
   \qquad 
   (\bar{\sigma}_{\mu\nu})_{\dot{\alpha}\dot{\gamma}}
 = \frac{i}{2}
  (\bar{\sigma}_{[\mu} )_{\dot{\alpha} \beta}
   ({\sigma}_{\nu ]} )_{\beta \dot{\gamma}}\,,
 \end{equation}
 and the complex generators $T_{ij}$ of $SU(N)$
 algebra satisfy the Hermicity condition
 \begin{equation}\label{mosluke3.7}
 T_{ij}= \bar{T}^{ji} \quad \Rightarrow \quad
 T_{ij} = T^{(S)}_{ij} + i\, T^{(A)}_{ij} \,,
 \end{equation}
 where $T^{(S)}_{ij}$ ($T^{(A)}_{ij})$ are real and symmetric (antisymmetric)
 \begin{equation}\label{mosluke3.8}
 T^{(S)}_{ij}= {T}^{(S)}_{ji}  \qquad
 T^{(A)}_{ij}= - {T}^{(A)}_{ji}  \qquad
 T^{(S)}_{ii}=0 \,.
 \end{equation}
 The trace of operator-valued matrix generators $T_{ij}$ has been separated and denoted by $A$ (axial charge).
 
 Under the D=4 conformal generators ($M_{\mu\nu}, P_\mu, K _\mu, D$) the supercharges $Q^{i}_{\alpha}, S^{i}_{\alpha}$ transform as follows
 \begin{equation}\label{mosluke3.9}
 \begin{array}{ll}
 [M_{\mu\nu}, Q_{\alpha i}]
 = - \frac{1}{2} (\sigma_{\mu\nu})_{\alpha}^{\ \beta} \, 
 Q_{\beta i}\,,
 \qquad\qquad
 &
   [ P_\mu , Q^{i}_{\alpha}] = 0 \,,
 \cr
 [M_{\mu\nu}, \bar{S}_{\dot{\alpha}}^{\ i}]
 = - \frac{1}{2} (\bar{\sigma}_{\mu\nu})_{\dot{\alpha}}^{\ \dot{\beta}} \, 
 \bar{S}_{\dot{\beta}}^{\ i}\,,
 \qquad  \qquad
 &
 [ P_\mu , \bar{S}^{i}_{\dot{\alpha}}] =  - (\bar{\sigma}_\mu)_{\dot{\alpha}}^{\ \dot{\beta}} \,
 \bar{S}^{i}_{\dot{\beta}}\,,
 \cr
 [K_{\mu}, {Q}_{{\alpha} i}]
 = -  ({\sigma}_{\mu})_{{\alpha}}^{\ \dot{\beta}} \, 
 \bar{S}_{\dot{\beta}}^{\ i}\,,
 \qquad \qquad &
 [ D, {Q}_{{\alpha} i}] =
   -\frac{i}{2} {Q}_{\alpha i}\,,
    \cr
 [K_{\mu}, \bar{S}_{{\alpha}}^{ i}]= 0\,,
 \qquad\qquad
 &
  [D, {\bar{S}}^{i}_{\dot{\alpha}}] = - \frac{i}{2} \bar{S}_{\dot{\alpha}}^{i} \,.
 \end{array}
 \end{equation}
 Further we obtain in consistency with the relation $T_{ii}=0$ that
 \begin{eqnarray}\label{mosluke3.10}
[ T_{ij}, Q_{\alpha k} ] = \delta_{ik} \, Q_{\alpha j}
 - \frac{1}{4} \, \delta_{ij}  Q_{\alpha k}\,,
 \cr\cr
 [T_{ij}, S_ {\alpha k} ] = \delta_{i k} \, S_{\alpha j}
 - \frac{1}{4} \, \delta_{j k}  S_{\alpha i}\,,
 \end{eqnarray}
 where $T_{ij}$ satisfy the $SU(N)$ algebra relations:
 \begin{equation}\label{mosluke3.11}
 [T_{ij}, T_{kl}] = \delta_{il} \, T_{jk} - \delta_{i k} \, T_{jl}\,.
 \end{equation}
 The $U(1)$ axial charge $A$ satisfies the relations
 \begin{eqnarray}\label{mosluke3.12}
 [A, Q_{\alpha k} ] = 
 \frac{i}{4}
 \left(
 1 - \frac{4}{N}
 \right) Q_{\alpha k}\,,
 \cr\cr
 [A, S_{\alpha k} ] = 
  \frac{i}{4}
 \left(
 1 - \frac{4}{N}
 \right) S_{\alpha k} \,,
 \end{eqnarray}
 and commutes with all other bosonic generators forming 
 $O(2,4)\oplus SU(N)$ algebra.
 We see from (\ref{mosluke3.13}) that for $N=4$ the axial charge $A$ becomes a central charge.
 
 We shall further assume that $N=2k$ and  rewrite the D=4 relativistic superconformal algebra in different fermionic basis
 \begin{eqnarray}\label{mosluke3.13}
 Q^{\pm }_{\alpha i} & = &
 \frac{1}{2}
 \left(
 Q_{\alpha i} \pm
  \epsilon_{\alpha \beta}
 \Omega_{ij} \, \bar{Q}_{\dot{\beta}}^{j}
 \right)\,,
 \cr\cr
 \bar{Q}^{\pm i}_{\dot{\alpha}} & = &
 \frac{1}{2}
 \left(
 \bar{Q}^{i}_{\dot{\alpha}} \pm \epsilon_{\dot{\alpha} \dot{\beta}}
 \Omega^{ij} \, {Q}_{\beta j}
 \right)\,,
 \end{eqnarray}
 where the real matrix $\Omega$ ($\Omega^{ij}\equiv (\Omega_{ij})^\star = \Omega_{ij})$ is a $2k\times 2k$ symplectic metric ($\Omega^2= -1$,  $\Omega^T=-\Omega$). We choose for simplicity that
 \begin{equation}\label{mosluke3.14}
 \Omega =
 \begin{pmatrix}
 0 & - \mathbb{1}_k \cr
 \mathbb{1}_k & 0
 \end{pmatrix}
 \end{equation}
 
 The Weyl supercharges (\ref{mosluke3.13}) satisfy the subsidiary symplectic-Majorana conditions 
 \begin{equation}\label{mosluke3.15}
 \bar{Q}^{\pm i}_{\ \dot{\alpha}}
 = \pm \epsilon_{\alpha\beta}
 \Omega^{ij}\, Q^{\pm }_{\beta j} \,.
 \end{equation}
 Similarly one can introduce the supercharges $S^{\pm i}_{\alpha}$, 
 $\bar{S}^{\pm i}_{\alpha}$, with the same subsidiary condition
 \begin{equation}\label{mosluke3.16}
 \bar{S}^{\pm i}_{\dot{\alpha}} = \pm \epsilon_{\alpha\beta} \, \Omega^{ij} \, S^{\pm }_{\ \beta j} \,.
 \end{equation}
 The relations (\ref{mosluke3.1}--\ref{mosluke3.4}) can be represented now as follows ($r=1,2,3$)
 \begin{eqnarray}\label{mosluke3.17}
 \{{Q}^{\pm }_{{\alpha} i} ,
 {\bar{Q}}^{\pm j}_{\dot{\beta}}
 \}
 & = & 
 \delta_{i}^{\, j} \, \delta_{{\alpha} \dot{\beta}} P_0 \,\,,
 \cr
  \{{Q}^{\pm }_{{\alpha} i},
  \bar{Q}^{\mp j}_{\dot{\beta}}
   \}
 & = & 
 \delta_{i}^{\,j} \,  (\sigma_r P_r)_{{\alpha} \dot{\beta}} \,\,,
 \cr
  \{ {Q}^{\pm }_{{\alpha} i},
  {Q}^{\pm }_{{\beta} j}
   \}
 &  = &
    \{ {Q}^{\pm }_{{\alpha} i},
  {Q}^{\mp }_{\dot{\beta} j}
   \}
   =
    \{ \bar{Q}^{\pm i }_{\dot{\alpha} },
  \bar{Q}^{\pm j}_{\dot{\beta} } 
   \}
   =
    \{ \bar{Q}^{\pm i}_{\dot{\alpha}},
  \bar{Q}^{\mp j}_{\dot{\beta}}
   \} 
  = 0 \,\,,
    \end{eqnarray}
 and
 \begin{eqnarray}\label{mosluke3.18}
 \{ {S}^{\pm i}_{{\alpha}},
  {\bar{S}}^{\pm j}_{\dot{\beta}} 
   \}
   & = & 
- \delta_{i}^{\, j} \,  \delta_{{\alpha} \dot{\beta}} K_0\,,
\cr
\{ {S}^{\pm i}_{{\alpha}},
  {\bar{S}}^{\mp j}_{\dot{\beta}} 
   \}
   & = & 
- \delta_{i}^{\, j} \,  (\sigma_r \, K_r)_{{\alpha} \dot{\beta}} \,,
\cr
\{ {S}^{\pm}_{{\alpha i}},
  {S}^{\pm }_{{\beta} j} 
   \}
   & = & 
   \{ {S}^{\pm }_{{\alpha} i},
  {S}^{\mp }_{{\beta} j} 
   \}
   = \{ \bar{S}^{\pm i} _{\dot{\alpha}}, \bar{S}^{\pm j} _{\dot{\beta}} \}
   = \{ \bar{S}^{\pm i} _{\dot{\alpha}}, {S}^{\mp j} _{\dot{\beta}} \}
   =0\,.
 \end{eqnarray}
 Further from (\ref{mosluke3.5a}--\ref{mosluke3.5d}), (\ref{mosluke3.6}) one obtains
 \begin{subequations}
\begin{eqnarray}\label{mosluke3.19}
\{Q^{\pm}_{\alpha i}, S^{\pm}_{\beta j} \}
&=& \frac{1}{2} \, \delta_{ij}
\big[
(\sigma_{rs})_{\alpha \beta}\, M_{rs}
- \epsilon_{\alpha \beta}\, D
\big]
+
\frac{1}{2}\epsilon_{\alpha \beta}
\left(
T^{(-)S}_{ij} + i\, T^{(+)A}_{ij}
\right)\,,
\label{mosluke3.19a}
\\[12pt]
\{Q^{+}_{\alpha i}, S^{-}_{\beta j} \}
&=& \frac{1}{2} \, \delta_{ij}
\big[
(\sigma_{r0})_{\alpha \beta}\, M_{r0}
- \epsilon_{\alpha \beta}\, A
\big]
-
\frac{1}{2}\epsilon_{\alpha \beta}
\left(
T^{(+)A}_{ij} - i\, T^{(+)S}_{ij}
\right)\,,
\label{mosluke3.19b}
\\[12pt]
\{ Q^{\pm}_{\alpha i}, \bar{S}^{\pm j}_{\dot{\beta}} \}
&=& \{ Q^{\pm}_{\alpha i} , \bar{S}^{\mp j}_{\dot{\beta}} \} =0\,,
\label{mosluke3.19c}
\end{eqnarray} 
\end{subequations}
and corresponding complex conjugate relations,  where 
 $T^{(S)}_{kj} = T^{+(S)}_{kj} + T^{+(S)}_{kj}$,
 $T^{(A)}_{kj} = T^{+(A)}_{kj}+T^{-(A)}_{kj}$
 and
 \begin{equation}\label{mosluke3.20}
 \Omega_{ik}T^{\pm(S)}_{kj}=
 \pm T^{\pm(S)}_{ik} \Omega_{kj}\,, \qquad \quad
  \Omega_{ik}T^{\pm(A)}_{kj}=
 \pm T^{\pm(A)}_{ik} \Omega_{kj} \,.
 \end{equation}
 The internal symmetry generators $T^{ij}$ do split in relations (\ref{mosluke3.19a}--\ref{mosluke3.19b})
 into the following two sets of $SU(2k)$ algebra generators
 \begin{eqnarray}\label{mosluke3.21}
 \Omega \cdot \mathbb{H} = - \mathbb{H}^T \, \Omega \,, &
 \qquad \qquad &
 \mathbb{H} = \left(T^{-(S)}_{ij}, T^{+(A)}_{ij}
 \right)\,,
 \cr
  \Omega \cdot \mathbb{K} = - \mathbb{K}^T \, \Omega\,, &
 \qquad \qquad &
 \mathbb{K} = \left(T^{+(S)}_{ij}, T^{-(A)}_{ij}
 \right)\,,
 \end{eqnarray}
 where
 \begin{equation}\label{mosluke3.22}
 \mathbb{H} = \textit{USp}(2k)\,,
 \qquad \qquad
 \mathbb{K} = \frac{\textit{SU}(2k)}{\textit{USp}(2k)}\,,
 \end{equation}
 provide an example of symmetric Riemannian space ($\mathbb{H},\mathbb{K}$) with the algebraic relations
 \begin{equation}\label{mosluke3.23}
 [\mathbb{H}, \mathbb{H}] \subset
 \mathbb{H} \qquad
 [\mathbb{H}, \mathbb{K}] \subset
 \mathbb{K}
 \qquad
 [\mathbb{K}, \mathbb{K}] \subset
 \mathbb{H}\,.
 \end{equation}
 
 In order to obtain the Galilean conformal superalgebra (GCSA) we should introduce the rescaling
 \begin{eqnarray}\label{mosluke3.24}
 Q^{+}_{\alpha i} = \frac{1}{\sqrt{c}} \, \widetilde{Q}^{+}_{\alpha i}\,,
 \qquad \qquad 
  Q^{-}_{\alpha i} = {\sqrt{c}} \, \widetilde{Q}^{-}_{\alpha i}\,,
  \cr
   S^{+}_{\alpha i} = {\sqrt{c}} \, \widetilde{S}^{+}_{\alpha i}\,,
   \qquad \qquad
   S^{-}_{\alpha i} = (c)^{3/2}
   \widetilde{S}^{-}_{\alpha i} \,,
 \end{eqnarray}
 where in the limit $c \to \infty$ the tilde denotes Galilean supercharges.
 Further we rescale
 \begin{eqnarray}\label{mosluke3.25}
 h_{ij} = \widetilde{h}_{ij}\,,
 \qquad \qquad
 h_{ij} \in USp(2k)\,,
 \nonumber \\[4pt]
 k_{ij} = c \widetilde{k}_{ij}\,,
 \qquad \qquad
 k_{ij} \in \frac{SU(2k)}{USp(2k)}\,,
 \\
 A = c \widetilde{A}\,, \qquad \qquad
 A \in U(1)\, \,,
 \nonumber
 \end{eqnarray}
 and from (\ref{mosluke3.23}) one gets
 \begin{equation}\label{mosluke3.26}
 [\widetilde{\mathbb{K}}, \widetilde{\mathbb{K}}] \subset \frac{1}{c^2} \, \mathbb{H} 
{\xrightarrow[c\to\infty]{\phantom{sssslhh}}}
   0 \,.
 \end{equation}
 The generators $\widetilde{k}_{ij}$ are becoming Abelian and they describe a set of internal complex momenta.
 One gets the following structure of Galilean conformal internal symmetry algebra \cite{mosluk16,mosluk17}
 \begin{equation}\label{mosluke3.27}
 \widetilde{\mathbb{T}} = \widetilde{\mathbb{K}} \niplus \widetilde{\mathbb{H}}
 \qquad \qquad
 \begin{array}{l}
 \widetilde{\mathbb{H}}= \textit{USp}(2k)
 \cr
 \widetilde{\mathbb{K}}=T^{k(2k-1)} \ \hbox{(Abelian)} \,.
 \end{array} 
 \end{equation}
 
 Substituting (\ref{mosluke3.24}) in $N=2k$ conformal superalgebra (\ref{mosluke3.18}--\ref{mosluke3.20}) and performing the limit $c\to \infty$, we get the following set of GSCA relations ($r=1,2,3$)
 \\
 \textbf{i)} \textit{Fermionic bilinear relations}
 \begin{eqnarray}\label{mosluke3.28}
 &\{
 \widetilde{Q}^{+}_{\alpha i},
 {\bar{\widetilde{Q}}}^{+  j}_{\dot{\beta}}
 \}
 = \delta_{i}^{\, j}\, \delta_{\alpha \dot{\beta}} \, H\,,
  \nonumber\\[4pt]
  &\{
 \widetilde{Q}^{+}_{\alpha i},
 {\bar{\widetilde{Q}}}^{- j}_{\dot{\beta}}
 \}
 =
 - \delta_{i}^{ j} (\sigma_r P_r)_{\alpha \dot{\beta}}\,,
 \nonumber\\[4pt]
  &\{
 \widetilde{Q}^{-}_{\alpha i},
 {\bar{\widetilde{Q}}}^{- j}_{\dot{\beta}}
 \}
 =
 0\,,
 \\ \cr
 &\{
 \widetilde{S}^{+}_{\alpha i},
 {\bar{\widetilde{S}}}^{+ j}_{\dot{\beta} }
 \}
 = - \delta_{i}^{j}\, \delta_{\alpha \dot{\beta}} \, K\,,
 \nonumber\\[2pt]
 &\{
 \widetilde{S}^{+}_{\alpha i},
 {\bar{\widetilde{S}}}^{- j}_{\dot{\beta} }
 \}
 =
 - \delta_{i}^{j} (\sigma_r F_r)_{\alpha \dot{\beta}}\,,
 \nonumber\\[2pt]
 &\{
 \widetilde{S}^{-}_{\alpha i},
 {\bar{\widetilde{S}}}^{- j}_{\dot{\beta} }
 \}
 = 0\,,
 \label{mosluke3.29}
 \\[12pt]
 &\{
 \widetilde{Q}^{+}_{\alpha i},
 {{\widetilde{S}}}^{+}_{{\beta} j}
 \}
 =
 \frac{1}{2}
 \delta_{ij}
 [
 (\sigma_{rs})_{\alpha \beta}
 \, M_{rs} - \epsilon_{\alpha \beta} \, D]
 + \frac{1}{2} \epsilon_{\alpha \beta}
 \left(
 T^{(-)S}_{ij} + i \, T^{(+)A}_{ij}
 \right)\,,
 \label{mosluke3.30}
 \\[12pt]
 &\{
 \widetilde{Q}^{\pm}_{\alpha i},
 {{\widetilde{S}}}^{\mp}_{{\beta} j}
 \}
 =
 \frac{1}{2}
 \delta_{ij}
 [
 (\sigma_{rs})_{\alpha \beta}
 \, B_{r} - i \epsilon_{\alpha \beta} \, \widetilde{A}]
 - \frac{1}{2} \epsilon_{\alpha \beta}
 \left(
 \widetilde{T}^{(+)A}_{ij} - i \, \widetilde{T}^{(-)S}_{ij}
 \right)\,,
 \label{mosluke3.31}
 \\[12pt]
 &\{
 \widetilde{Q}^{-}_{\alpha i},
 {{\widetilde{S}}}^{-}_{{\beta} j}
 \}
 = 0 \,.
 \label{mosluke3.32}
 \end{eqnarray}
 To the relations (\ref{mosluke3.30}--\ref{mosluke3.32}) one should add complex-conjugated relations.
 
 \textbf{ii)} \textit{Mixed bosonic-fermionic sector.}
 
 By rescaling (\ref{mosluke3.9}) one gets (we present only nontrivial commutators, the  complex-conjugate relations should be also added)
 
\begin{eqnarray}\label{mosluk3.33}
 &&[
 H, \widetilde{S}^{\pm }_{\alpha i}]
 = \widetilde{Q}^{\pm}_{\alpha i}\,,
 \qquad
 [P_r, \widetilde{S}^{+}_{\alpha i}]
 = - (\sigma_r)_{\alpha \dot{\beta}} \, 
 \bar{\widetilde{Q}}{}^{- i}_{\dot{\beta} }\,,
  \nonumber\\[2pt]
  &&[
 K, \widetilde{Q}^{\pm }_{\alpha i}]
 = \widetilde{S}^{\pm}_{\alpha i}\,,
 \qquad
 [F_r, \widetilde{Q}^{-}_{\alpha i}]
 = - (\sigma_r)_{\alpha \dot{\beta}} \, 
 \bar{\widetilde{S}}{}^{+ i}_{\dot{\beta} }\,,
 \\[8pt]
 \label{mosluke3.34}
 &&[M_{ij}, \widetilde{Q}^{\pm}_{\alpha i}]
 = - \frac{1}{2} (\sigma_{ij})_{\alpha {\beta}} \, 
 {\widetilde{Q}}{}^{\pm}_{{\beta} i}\,,
 \qquad 
 [M_{ij}, \widetilde{S}^{\pm}_{\alpha i}]
 = - \frac{1}{2} (\sigma_{ij})_{\alpha {\beta}} \, 
 {\widetilde{S}}{}^{\pm}_{{\beta} i}\,,
  \nonumber\\[2pt]
 &&[B_{r}, \widetilde{Q}^{+}_{\alpha i}]
 = - \frac{1}{2} (\sigma_{or})_{\alpha {\beta}} \, 
 {\widetilde{Q}}{}^{-}_{{\beta} i}\,,
 \qquad 
 [B_{r}, \widetilde{S}^{-}_{\alpha i}]
 = - \frac{1}{2} (\sigma_{or})_{\alpha {\beta}} \, 
 {\widetilde{S}}{}^{+}_{{\beta} i}\,,
 \\[8pt]
 \label{mosluk3.35}
 &&[D, \widetilde{Q}^{\pm}_{\alpha i} ]
 = - \frac{1}{2} \widetilde{Q}^{\pm}_{\alpha i}\,,
 \qquad
 [D, \widetilde{S}^{\pm}_{\alpha i} ]
 = \frac{1}{2} \widetilde{S}^{\pm}_{\alpha i}\,,
  \nonumber\\[2pt]
 &&[A, \widetilde{Q}^{+}_{\alpha i} ]
 = \frac{i}{4} (1 - \frac{4}{N}) Q^{-}_{\alpha i}\,,
 \qquad
 [A, \widetilde{S}^{+}_{\alpha i} ]
 = \frac{i}{4} (1 - \frac{4}{N}) \widetilde{S}^{-}_{\alpha i}\,.
 \end{eqnarray}
 
 The covariance relations with respect to the generators $T_{ij}=h_{ij}\oplus \widetilde{k}_{ij}$ are the following
 
 1) $h_{ij}=(\widetilde{T}^{-(S)}_{ij}, \widetilde{T}^{+(A)}_{ij}) \in USp(2k)$
 
\begin{eqnarray}\label{mosluke3.36}
 &&
 \left[T^{-(S)}_{ij}, \widetilde{Q}^{\pm}_{\alpha k}\right]
 = - i \left(\tau^{-(S)}_{ij}\right)_{kl}\, \widetilde{Q}^{\pm}_{\alpha l}\,,
 \nonumber\\[8pt]
 &&
 \left[T^{+(A)}_{ij}, \widetilde{Q}^{\pm}_{\alpha k}\right]
 =  \left(\tau^{+(A)}_{ij}\right)_{kl}\, \widetilde{Q}^{\pm}_{\alpha l}\,,
  \nonumber\\[8pt]
 &&
 \left[T^{-(S)}_{ij}, \widetilde{S}^{\pm}_{\alpha k}\right]
 = - i \left(\tau^{(-)S}_{ij}\right)_{kl}\, \widetilde{S}^{\pm}_{\alpha l}\,,
  \nonumber\\[8pt]
 &&
 \left[T^{+(A)}_{ij}, \widetilde{S}^{\pm}_{\alpha k}\right]
 =   \left(\tau^{+(A)}_{ij}\right)_{kl}\, \widetilde{S}^{\pm}_{\alpha l} \ .
 \end{eqnarray}

  2) $\widetilde{k}_{ij}=({T}^{+(S)}_{ij}, {T}^{-(A)}_{ij}) \in T^{k(2k-1)}$
 
 \begin{eqnarray}\label{mosluke3.37}
 &&
 \left[T^{+(S)}_{ij}, \widetilde{Q}^{+}_{\alpha k} \right]
 = - i (\tau^{+(S)}_{ij})_{kl}\, \widetilde{Q}^{-}_{\alpha l}\,,
 \nonumber\\[4pt]
 &&
 \left[T^{-(A)}_{ij} , \widetilde{Q}^{+}_{\alpha k}\right]
 =  (\tau^{-(A)}_{ij})_{kl}\, \widetilde{Q}^{-}_{\alpha l}\,,
 \nonumber\\[4pt]
 &&
 \left[T^{+(S)}_{ij}, \widetilde{S}^{+}_{\alpha k}\right]
 = - i (\tau^{+(S)}_{ij})_{kl}\, \widetilde{S}^{-}_{\alpha l}\,,
 \nonumber\\[4pt]
 &&
 \left[T^{-(A)}_{ij}, \widetilde{S}^{+}_{\alpha k}\right]
 =   (\tau^{-(A)}_{ij})_{kl}\, \widetilde{S}^{-}_{\alpha l} \,.
 \end{eqnarray}
 
 It should be stressed that the generators $\widetilde{k}_{ij}$ commute with the supercharges $Q^{-}_{\alpha i}, S^{-}_{\alpha i}$, and we note that $(\tau^{?}_{ij})_{kl}$ are the $2k\times 2k$
 matrix realizations of the generators $T^{?}_{ij}$.
 
 The relations (\ref{mosluke3.28}--\ref{mosluke3.29}) describe the supersymmetric extension of the bosonic sector spanned by the summ of the GCA generators  ($M_{ij},H,D,K_l,P_l,B_l,F_i$) and the internal $U(N)$ sector
 ($T^{ij}, A$).
 As well follows from relations (\ref{mosluke3.37})  that the Abelian generators $\widetilde{k}_{ij}$ should not be treated as internal tensorial central charges.
 
 We would like add the following three comments:
 
 \textbf{i)} The rescaling (\ref{mosluke3.24}--\ref{mosluke3.25}) is the extension of the ``natural'' NR rescaling (\ref{mosluke2.1}) to $SU(2,2|2k)$
  which leads to D=4 extended GSCA \cite{mosluk16}.
  It is possible however to introduce simpler rescaling of $SU(2,2|2k)$ by generalizing the formula (\ref{mosluke2.5}) \cite{mosluk17}. 
  Let us introduce the following $Z_3$-grading of $SU(2,2|2k)$ superalgebra
  \begin{equation}\label{mosluke3.38}
  \begin{array}{ccc}
  L_0 &L_1 &L_2
  \cr
  M_{ij}\oplus (P_0,K_0,D)\oplus h_{ij}\qquad \qquad
  &
  Q_{\alpha i}\oplus Q_{\dot{\alpha} i} \oplus
  S_{\alpha i} \oplus \bar{S}_{\dot{\alpha} i}
  \qquad \qquad
  & M_{0i}\oplus P_i \oplus K_i \oplus K_{ij}
  \end{array}
  \end{equation} 
  corresponding to the following coset product structure of
   $SU(2,2;2k)$
   \begin{equation}\label{mosluke3.39}
   SU(2,2;2k) = H \cdot K \cdot F \,\,,
   \end{equation}
   where
   \begin{eqnarray}\label{mosluke3.40}
   &&
   \mathbb{H} = SO(3)\otimes SO(2,1) \otimes USp(2k)\,,
   \nonumber\\[4pt]
   && 
   K = \frac{SO(4,2)\otimes U(N)}{SO(3)\otimes SO(2,1)\otimes USp(2k)}\,,
    \nonumber\\[4pt]
   &&
   F = \frac{SU(2,2|2k)}{SO(4,)\otimes U(N)}\,.
   \end{eqnarray}
 Denoting the corresponding generators  in the decomposition (\ref{mosluke3.39}) by $h_I , k_J , f_\rho$, one can introduce the rescaled generators
 \begin{equation}\label{mosluke3.41}
 h'_I =h_I \qquad \quad
 k'_J = \lambda\, k_J \qquad \quad
 f'_\rho = \lambda^{\frac{1}{2}} \, f_\rho \,,
 \end{equation}
 which extends from $O(4,2)$ to $SU(2,2|2k)$
  the rescaling (\ref{mosluke2.5}). It can be checked that in the contraction limit
  $\lambda \to \infty$ one obtains the same
    GCSA as derived by natural rescaling (\ref{mosluke3.24}--\ref{mosluke3.25}) \cite{mosluk17} (see also \cite{mosluk11}).
 
 \textbf{ii)} The relations of GSCA has been obtained in \cite{mosluk16,mosluk17}. Almost at the same time did appear the paper \cite{mosluk18} proposing other NR contraction scheme and alternative supersymmetrization of GCA. In the framework proposed in \cite{mosluk18}  the contraction could be performed for $SU(2,2|N)$ for any $N$. In particular for $N=1$ there was obtained the variant of GSCA with fermionic algebraic relations  expressing the generators $H, K, D$ and $M_i$ as bilinears in supercharges.
 It appears  that in \cite{mosluk18}  only the Abelian part of CGA generators is supersymmetrized, and in particular the basic relation of SUSY QM ($\{ Q, Q\} \sim H$) is not included among the proposed  relations in fermionic sector. We conclude therefore that the supersymmetrization of GCA presented in \cite{mosluk18}  from the physical point of view is not an attractive one. 
 
 \textbf{iii)} The GCA has a quaternionic structure. Indeed, the stability subgroup $H$ contains the quaternionic groups $SU(2)\simeq U(1;H)$ and $USp(2k)\simeq U(k;H)$, i.e. the Galilean conformal supercharges can be described as the pairs of $k$-dimensional quaternionic supercharges. Let us observe that two-component $SU(2)$ spinor $Z_\alpha =(z_1,z_2)$ can be described by a single quaternion $q$ (see e.g. \cite{mosluk30})
 \begin{equation}\label{mosluke3.42}
 (z_1=x_1 + iy_1, z_2 = x_2 + i y_2)
 \leftrightarrow q^H = x_1 +e_3 y_1 +
 e_2(x_2 + e_3 y_2) \,.
 \end{equation} 
 From (\ref{mosluke3.16}--\ref{mosluke3.17}) follows that only the complex charges $\widetilde{Q}^{\pm }_{\alpha a}$, 
  $\widehat{Q}^{\pm }_{\dot{\alpha} a}$,
  $\bar{S}^{\pm }_{\alpha a}, {\bar{\widetilde{S}}}_{\dot{\alpha} a}$ 
   ($a=1 \ldots k$) are independent, i.e. GCSA can be formulated in terms of the pair of $k$-dimensional quaternionic supercharges
   $\widetilde{Q}^{\pm H}_{a}$, 
    $\widetilde{S}^{\pm H}_{a}$ where the indices $a$ span the quaternionic representation space of Galilean compact  internal symmetry $U(k;H)$ group. It is a matter of quite tedious calculations (compare e.g. with \cite{mosluk30}, Sect.~6) to reexpress GSCA in the quaternionic form.
    
    \section{Final Remarks}
 
 In this paper we briefly reviewed the description of Galilean conformal algebra and its supersymmetrization. These recently described new symmetries and supersymmetries can be applied to the nonrelativistic version of AdS/CFT correspondence or to the geometric description of nonrelativistic counterparts of conformal gravity and conformal supergravity. We add also that we  did not consider here the Newton-Hooke symmetries, describing nonrelativistic space-times with constant spatial curvature (nonrelativistic counterparts of AdS and dS geometries) \cite{mosluk31}--\cite{mosluk35} as well as their supersymmetrization \cite{mosluk11,mosluk17}.
 
\begin{acknowledgments}
 The author would like to thank J.A.~de~Azcarraga, S.~Fedoruk, E.~Ivanov, P.~Kosi\'{n}ski and P.~Ma\'{s}lanka for fruitful  collaboration on the subjects described in this talk, and 
  G.~Pogosyan for his warm hospitality in Armenia.
 It is also acknowledged  the support from  Polish Ministry of Science and Higher Education by grant N~202331139.
\end{acknowledgments}

\end{document}